\documentclass[preprint,12pt]{elsarticle}

\usepackage{lipsum}

\usepackage{lineno}

\usepackage{amsmath,lscape}
\usepackage{amssymb}
\usepackage[utf8]{inputenc}

\usepackage{epsfig}
\usepackage{amsfonts}
\usepackage{fullpage}
\usepackage{afterpage}
\usepackage{rotating,array,booktabs} 
\usepackage{enumerate}
\usepackage[shortlabels]{enumitem}
\usepackage{wrapfig}
\usepackage{subfig}

\usepackage{multirow,multicol}
\usepackage{siunitx}
\usepackage{pdflscape}
\usepackage{afterpage}
\usepackage[nolist]{acronym}
\usepackage{longtable}

\usepackage{algpseudocode}
\usepackage{algorithm}

\usepackage{hyperref}
\hypersetup{
    unicode=true,          
    pdftoolbar=true,        
    pdfmenubar=true,        
    pdffitwindow=false,     
    pdfstartview={FitH},    
    pdfauthor={M\'arcia de F\'atima Morais, Matheus Henrique Dal Molin Ribeiro, Ramon Gomes da Silva, Leandro dos Santos Coelho},     
    pdfnewwindow=true,      
    colorlinks=true,        
    linkcolor= black,        
    citecolor= black,        
    filecolor= black,        
    urlcolor= black          
    breaklinks=true
}

\usepackage[table,xcdraw]{xcolor}

\begin{document}
    
    \begin{frontmatter}
        \title{Short-term forecasting COVID-19 cumulative confirmed cases: Perspectives for Brazil}

\author[ad1,ad2]{Matheus Henrique Dal Molin Ribeiro}\corref{mycorrespondingauthor}
\ead{matheus.dalmolinribeiro@gmail.com}
\cortext[mycorrespondingauthor]{Corresponding author}

\author[ad1]{Ramon Gomes da Silva}

\author[ad3,ad4]{Viviana Cocco Mariani}

\author[ad1,ad4]{Leandro dos Santos Coelho}

\address[ad1]{Industrial \& Systems Engineering Graduate Program (PPGEPS), Pontifical Catholic University of Parana (PUCPR). 1155, Rua Imaculada Conceicao, Curitiba, PR, Brazil. 80215-901}

\address[ad2]{Department of Mathematics, Federal Technological University of Parana (UTFPR). Via do Conhecimento, KM 01 - Fraron,  Pato Branco, PR, Brazil. 85503--390}

\address[ad3]{Mechanical Engineering Graduate Program (PPGEM), Pontifical Catholic University of Parana (PUCPR). 1155, Rua Imaculada Conceicao, Curitiba, PR, Brazil. 80215-901}

\address[ad4]{Department of Electrical Engineering, Federal University of Parana (UFPR). 100, Avenida Coronel Francisco Heraclito dos Santos, Curitiba, PR, Brazil. 81530-000}

\journal{Chaos, Solitons \& Fractals}
        \begin{abstract}
The new Coronavirus (COVID-19) is an emerging disease responsible for infecting millions of people since the first notification until nowadays. Developing efficient short-term forecasting models allow knowing the number of future cases. In this context, it is possible to develop strategic planning in the public health system to avoid deaths. In this paper, autoregressive integrated moving average (ARIMA), cubist (CUBIST), random forest (RF), ridge regression (RIDGE), support vector regression (SVR), and stacking-ensemble learning are evaluated in the task of time series forecasting with one, three, and six-days ahead the COVID-19 cumulative confirmed cases in ten Brazilian states with a high daily incidence. In the stacking learning approach, the cubist, RF, RIDGE, and SVR models are adopted as base-learners and Gaussian process (GP) as meta-learner. The models' effectiveness is evaluated based on the improvement index, mean absolute error, and symmetric mean absolute percentage error criteria. In most of the cases, the SVR and stacking ensemble learning reach a better performance regarding adopted criteria than compared models. In general, the developed models can generate accurate forecasting, achieving errors in a range of 0.87\% - 3.51\%, 1.02\% - 5.63\%, and 0.95\% - 6.90\% in one, three, and six-days-ahead, respectively. The ranking of models in all scenarios is SVR, stacking ensemble learning, ARIMA, CUBIST, RIDGE, and RF models. The use of evaluated models is recommended to forecasting and monitor the ongoing growth of COVID-19 cases, once these models can assist the managers in the decision-making support systems.
\end{abstract}

\begin{keyword}
ARIMA \sep COVID-19 \sep Forecasting \sep Decision-making \sep Machine learning  \sep Time-series
\end{keyword}
    \end{frontmatter}
    

    \section{Introduction \label{INT}}

The new Coronavirus (COVID-19) is an emerging disease responsible for infecting millions of people and killing thousands worldwide since the first notification until nowadays, according to the World Health Organization (WHO) \cite{whoCOVID19,SOHRABI202071}. Also according to WHO, Brazil registered 40.581 confirmed cases until April 22th 2020, holding the 12th position in the world ranking in the number of confirmed cases of COVID-19, and 2nd position in the Americas (behind the United States of America). 

Due to the impacts of the COVID-19 pandemic in people’s lives and the world’s economy, the governments and population are most concerned with (i) when the COVID-19 outbreak will peak; (ii) how long the outbreak will last and (iii) how many people will eventually be infected \cite{zhang2020predicting}. Further, Boccaletti et al. \cite{boccaletti2020modeling} have identified at least three scientific communities that may cooperate in the effort to deal with the current pandemic: (i) the community of applied mathematicians, virologists and epidemiologists, developing sophisticated diffusion models to the specific properties of a given pathogen; (ii) the community of complex systems scientists who study the spread of infections using compartmental models, using methods and principles from statistical mechanics and nonlinear dynamics; and (iii) the community of scientists who incorporate artificial intelligence (AI) and most specifically deep learning approaches to produce accurate predictive models. Also, different studies are evaluating the impacts of COVID-19 on society, whether through predictions of future cases, as well as variables capable of helping to understand the spread of this disease \cite{becerra2020forecasting, fanelli2020analysis,FONG2020106282,ROOSA2020256,EFFENBERGER2020}.

Moreover, epidemiological time series forecasting plays an important role in health public system, once it allows the managers to develop strategic planning to avoid possible epidemics. Forecasting diseases as accurate as possible is important due to their impact on the public health system. To ensure this accuracy, AI models have been widely used to forecast epidemiological time series over the years \cite{davis2019genetic, ribeiro2019forecasting,SCAVUZZO2018167}. Moreover, in the AI context, Vaishya et al. \cite{VAISHYA2020337} presented a review of trends in COVID-19 data analysis.

Regarding this context, the objective of this paper is to explore and compare the predictive capacity of machine learning regression and statistical models, in the task of forecasting one, three, and six-days-ahead COVID-19 cumulative cases in Brazil. In this respect, datasets of ten Brazilian states some with a high incidence of COVID-19 until now, like Sao Paulo and Rio de Janeiro, are adopted to evaluates the forecasting efficiency through of the autoregressive integrated moving average (ARIMA), cubist regression (CUBIST), random forest (RF), ridge regression (RIDGE), support vector regression (SVR), and stacking-ensemble learning models. In the stacking learning modelling, which is an effective ensemble learning approach \cite{ribeiro2020ensemble, moreno2019very}, CUBIST, RF, RIDGE, and SVR are used as base-learners (weak models), and Gaussian process (GP) as meta-learner (strong model). The out-of-sample forecasting accuracy of each model is compared by some performance metrics such as the improvement percentage index (IP), mean absolute errors (MAE), and symmetric mean absolute percentage error (sMAPE).

The contributions of this paper can be summarized as follows: 

\begin{itemize}
    \item The first contribution is related to the presentation of a novel analysis of the forecast model for cumulative confirmed cases of COVID-19 in Brazil, whose accuracy of the models assists governors in decision-making to contain the pandemic and strategies concerning the health system; 
    \item The second contribution, we can highlight the use of heterogeneous machine learning models, as well as the stacking-ensemble learning approach to forecast the Brazilian cumulative confirmed cases of COVID-19; 
    \item Also, this paper evaluates models forecasting in a multi-day-ahead forecasting strategy. The forecasting time horizons are the interval of one, three, and six-days-ahead. This range of the forecasting time horizon allows us to verify the effectiveness of the predicting models in different scenarios, helping in future strategies in fighting COVID-19.
\end{itemize}

The remainder of this paper is organized as follows: Section~\ref{material} a brief description of the dataset adopted in this paper is given. The forecasting models applied in this study are described in Section~\ref{sub:BL}. Section~\ref{MET} details the procedures applied in the research methodology. Results obtained and related discussion about models’ forecasting performance are given on Section~\ref{RES}. Finally, Section~\ref{CONC} concludes this work with considerations and some directions for future research proposals.
    \section{Material and Methods \label{MAM}}

This section presents the description of the material analyzed (Section \ref{material}) as well as the models description applied in this paper (Section \ref{sub:BL}).

\subsection{Dataset Description \label{material}}

\begin{figure}[htb!]
    \centering
    \includegraphics[width=.6\linewidth]{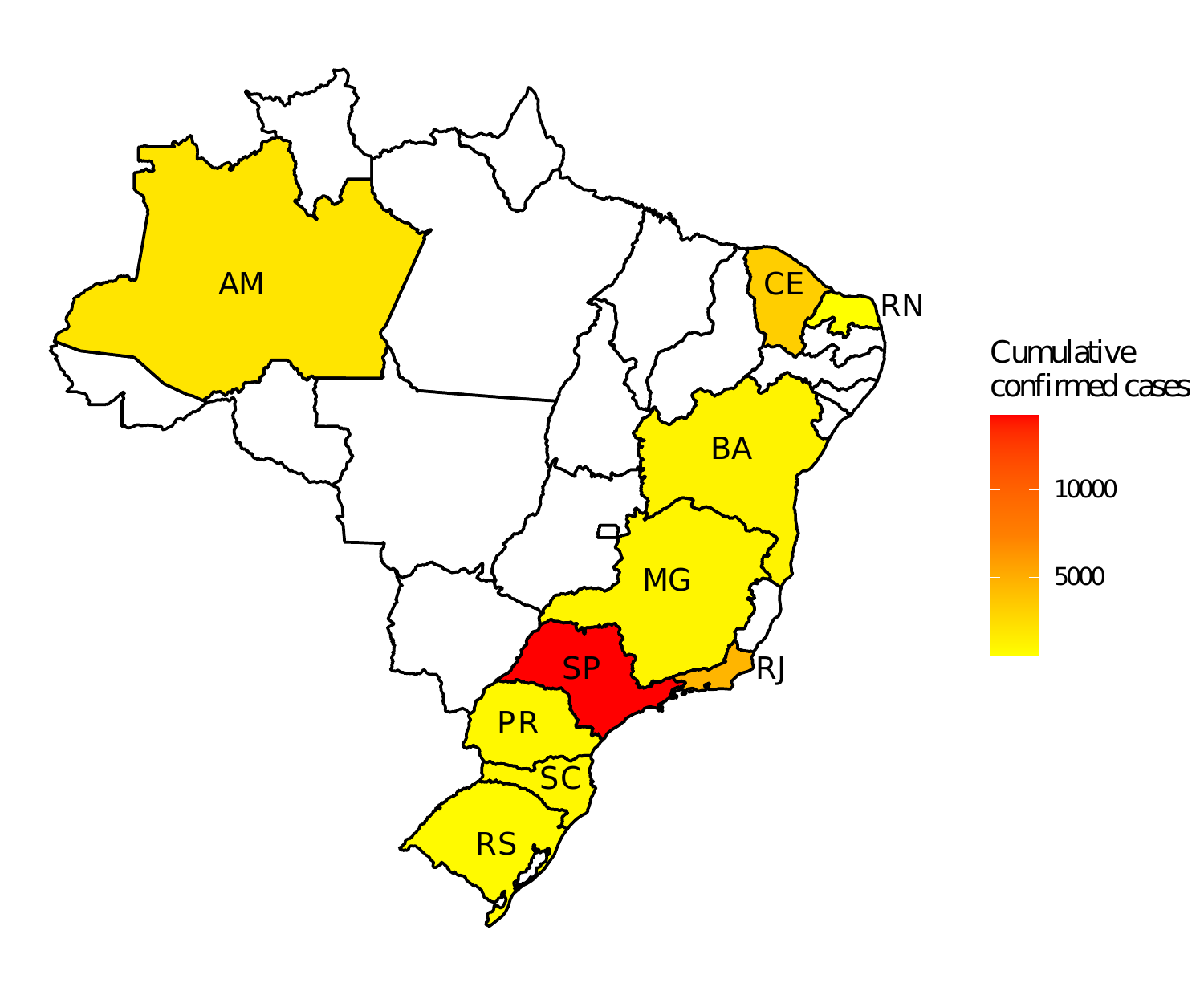}
    \caption{Heatmap of the cumulative confirmed cases of the analyzed states}
    \label{fig:heatmap}
\end{figure}

The collected dataset refers to the cumulative confirmed cases of COVID-19 that occurred in Brazil until April, 18 or 19 of 2020. The dataset was collected from an API \cite{brasilio} that retrieves the daily information about COVID-19 cases from all 27 Brazilian State Health Offices, gather them, and make it a publicly available. Among the 27 federative units (26 states and one federal district), ten states some with a high incidence of COVID-19 cases and other states with lower temperatures, states from south of Brazil, were chosen, among them are Amazonas (AM), Bahia (BA), Ceara (CE), Minas Gerais (MG), Parana (PR), Rio de Janeiro (RJ), Rio Grande do Norte (RN), Rio Grande do Sul (RS), Santa Catarina (SC), and Sao Paulo (SP). The measurement period of each state varies, once each state counts since the day of its first case until the day of the last report. The cumulative confirmed cases and deaths of each state, as well as the period from the first and last reports, are illustrated in Table~\ref{tab:reports}. The change in the way of accounting for the number of cases, by the health departments, may change the data presented here.
\begin{table}[htb!]
\centering
\caption{First and last report dates by state}
\label{tab:reports}
\begin{tabular}{cccccc}
  \hline
  State &
  \begin{tabular}[c]{@{}c@{}}Number of\\ observed days\end{tabular} & 
  First report & 
  Last report &
  \begin{tabular}[c]{@{}c@{}}Cumulative\\ confirmed cases\end{tabular} &
  \begin{tabular}[c]{@{}c@{}}Cumulative\\ deaths\end{tabular} \\ 
  \hline
 AM & 34 & 13/03/2020 & 19/04/2020 & 2044 & 182 \\ 
  BA & 43 & 06/03/2020 & 19/04/2020 & 1249 & 45 \\ 
  CE & 35 & 16/03/2020 & 19/04/2020 & 3306 & 189 \\ 
  MG & 42 & 08/03/2020 & 19/04/2020 & 1154 & 39 \\ 
  PR & 36 & 12/03/2020 & 18/04/2020 & 960 & 49 \\ 
  RJ & 38 & 05/03/2020 & 19/04/2020 & 4675 & 402 \\ 
  RN & 30 & 12/03/2020 & 18/04/2020 & 561 & 26 \\ 
  RS & 38 & 10/03/2020 & 19/04/2020 & 869 & 26 \\ 
  SC & 39 & 12/03/2020 & 19/04/2020 & 1025 & 35 \\ 
  SP & 53 & 25/02/2020 & 19/04/2020 & 14267 & 1015 \\ 
   \hline
\end{tabular}
\end{table}
\newpage
A heatmap of the cumulative confirmed cases is presented in Figure~\ref{fig:heatmap}.

\subsection{Methodologies \label{sub:BL}}

This section describes a brief of each model employed in the data analysis. 

\begin{itemize}

     \item ARIMA is a Box \& Jenkins modelling usually employed to deal with non-stationary time series. In fact, the ARIMA model is full specified by autoregressive (\textit{p}), different degrees of trend differences (\textit{d}), and moving average operators (\textit{q}). These parameters are used do define the model order, and usually defined by grid-search, as well as by autocorrelation and partial autocorrelation function. In this context, the model is described as ARIMA\textit{(p,d,q)} \cite{box2015time}.
    
    \item CUBIST is a rule-based model, which performs predictions following the regression of trees principle \cite{Quinlan1993}. Through the use of a committee of the rules, and using the neighborhood concept similar to $k$-nearest-neighbor modelling, the final forecasting is obtained.
    
    \item GP is composed of a set of random variables Gaussian distributed and fully specified by its mean and covariance (kernel) function \citep{Rasmussen2004}. In this paper, the GP with a linear kernel is adopted.
    
    \item RIDGE is a regularized regression approach \cite{HOERL1970} which employs a penalization term in the ordinary least squares algorithm. It is an effective tool, once it reduces the bias of parameter estimates by controlling the standard errors. Moreover, the model can deal with inputs multi-collinearity problem. 
    
    \item RF is a bagging ensemble-based model, which combines the bagging advantages characterized by the creation of multiple samples, with refitting through of the bootstrap technique, from the same set of data, and random selection of predictors to compose each node of the decision tree \cite{breiman2001random}. RF is a fast and robust supervised learning method able to deal with the randomness of the time series. Furthermore, it is interesting because, in addition to being an ensemble approach, only the number of predictors for each node needs to be tuned.
    
    \item SVR consists in determining support vectors (points) close to a hyperplane that maximizes the margin between two-point classes obtained from the difference between the target value and a threshold. To deal with non-linear problems SVR takes into account kernel functions, which calculates the similarity between two observations. In this paper, the linear kernel is adopted. The main advantages of the use of SVR lies in its capacity to capture the predictor non-linearity and then use it to improve the forecasting cases. In the same direction, it is advantageous to employ this perspective in this case study adopted, since that the samples are small \cite{Drucker1997}.
    
    \item Stacked Generalization or stacking is an ensemble-based approach \cite{wolpert1992stacked} which combines through a meta-learner the predictions of a set of weak models (base-learners) to obtain a stronger learner. This approach usually operates into two levels, where in the first level the base-learners are trained and its predictions are obtained. In the next stage, a meta-learner uses, as inputs, the predictions of the previous level in the training phase. The stacking predictions are obtained from meta-learner. The main advantage of the stacking ensemble is that this approach can improve the accuracy and additionally reduce error variance \cite{ribeiro2020ensemble}. 
\end{itemize}

    \section{Proposed forecasting framework \label{MET}}

This section describes the main steps in the data analysis adopted by CUBIST, RF, RIDGE, SVR, and stacking models. Also, the ARIMA modelling is described.

    \textbf{Step 1}: \label{step1} Firstly, the raw data is split into training and test datasets. The test dataset is composed of six last observations, and the training dataset by the remain samples \cite{ribeiro2020ensemble}. The training data are centered by its mean value and divided by its standard deviation. To develops multi-days-ahead COVID-19 cases forecasting, recursive strategy is employed \cite{RODRIGUESMORENO2020112869}. In this aspect, one model is fitted for one-day-ahead forecasting. Next, the recursive strategy uses the forecasting value as an input for the same model to forecast the next step, continuing this manner until reaching the desirable horizon. The training structure adopted in this paper is stated as follows,

      \begin{align}
        y_{(t+1)} = f\left\{y_t,\ldots,y_{t+1-n_y}\right\} + \epsilon \quad \epsilon \sim N(\mathbf{0},\sigma^2),
      \end{align}
   
    \noindent in which $f$ is a function related to the adopted model in the training stage, $y_{t+1}$ is the COVID-19 case one-day-ahead, $n_y = 5$ are the past confirmed cases, $\epsilon$ is the random error, following a normal distribution with zero mean and constant variance. In this paper, the aim is to obtain the cases up to $H$ next days, especially up to 1 (ODA, one-day-ahead), 3 (TDA, three-days-ahead), and 6-days-ahead (SDA, six-days-ahead), respectively. The following structures are considered,
   
  \begin{equation}
   \hat{y}_{t+h}  =
     \begin{cases}
      f\left[y_{t},y_{t-1},\ldots,y_t,\ldots,y_{t-n_y+1}\right] & \text{if } h =1 \\
      f\left[\hat{y}_{t+h-1},\ldots,\hat{y}_{t+h},\ldots,y_t,\ldots,y_{t+h-n_y}\right] & \text{if } h \in [2,n_y] \\
      f\left[\hat{y}_{t+h-1},\ldots,\hat{y}_{t+h-n_y}\right] & \text{if } h \in [n_y+1,\ldots,H],
     \end{cases}
  \end{equation}
    
   \noindent where $\hat{y}_{t+h}$ is the forecast value at time $t$ and forecast horizon up to $h$, $y_{t+h-n_y}$ and $\hat{y}_{t+h-n_y}$ are the previously observed and forecast cases lags in $n_y = 5$ days. The $n_y$ value is chosen through grid-search with purpose to capture the best data behavior. 
   
  \textbf{Step 2}: In the stacking modelling, the base-learners CUBIST, RF, RIDGE, SVR are trained and its forecasting are used as inputs for meta-learner GP. In the training stage, leave-one-out cross-validation with a time slice is adopted \cite{ribeiro2020ensemble}.  Finally, the out-of-sample forecasts are computed. These approaches are developed using the \texttt{caret} package \cite{caret2017}. The ARIMA modeling is performed through the use of \texttt{forecast} package \cite{forecastpackage,Rob2008} with use of \texttt{auto.arima} function. To define the ARIMA order, grid-search is adopted, and the most suitable order is that reach a lower Akaike and Bayesian Akaike criteria information. Both analyses are developed using \texttt{R} software \cite{R}. All hyperparameters employed in this study are presented in Table \ref{tab:hyper} in \ref{apendixB}. 
    
  \textbf{Step 3}: To evaluate the effectiveness of adopted models, from obtained forecasts out-of-sample (test set), performance IP \eqref{eq:criteria1}, MAE \eqref{eq:criteria2}, and sMAPE \eqref{eq:criteria3} criteria are computed as 

    \begin{equation}
        \begin{aligned}
            \operatorname{MAE} &=&\displaystyle \frac{1}{n} \sum_{i=1}^{n} \left|y_i-\hat{y}_i\right| ,      \label{eq:criteria1}
        \end{aligned}
    \end{equation}  
            
    \begin{equation}
        \begin{aligned}
            \operatorname{sMAPE}&=&\frac{1}{n}\displaystyle \sum_{i=1}^{n} \left|\frac{\hat{y}_i-y_i}{(|y_i|+|\hat{y}_i|/2)}\right|,
                  \label{eq:criteria2}
        \end{aligned}
    \end{equation}  
 \begin{equation}
        \begin{aligned}
            \operatorname{IP}&=&100 \times \frac{\displaystyle M_c-M_b}{M_c},
        \label{eq:criteria3}
        \end{aligned}
    \end{equation}  
    
  \noindent where $n$ is the number of observations, $y_i$ and $\hat{y}_{i}$ are the \textit{i}-th observed and predicted values, respectively. Also, the $M_c$ and $M_b$ represent the performance measure of compared and best models, respectively.

\begin{figure}[htb!]
    \centering
    \includegraphics[width=0.8\linewidth]{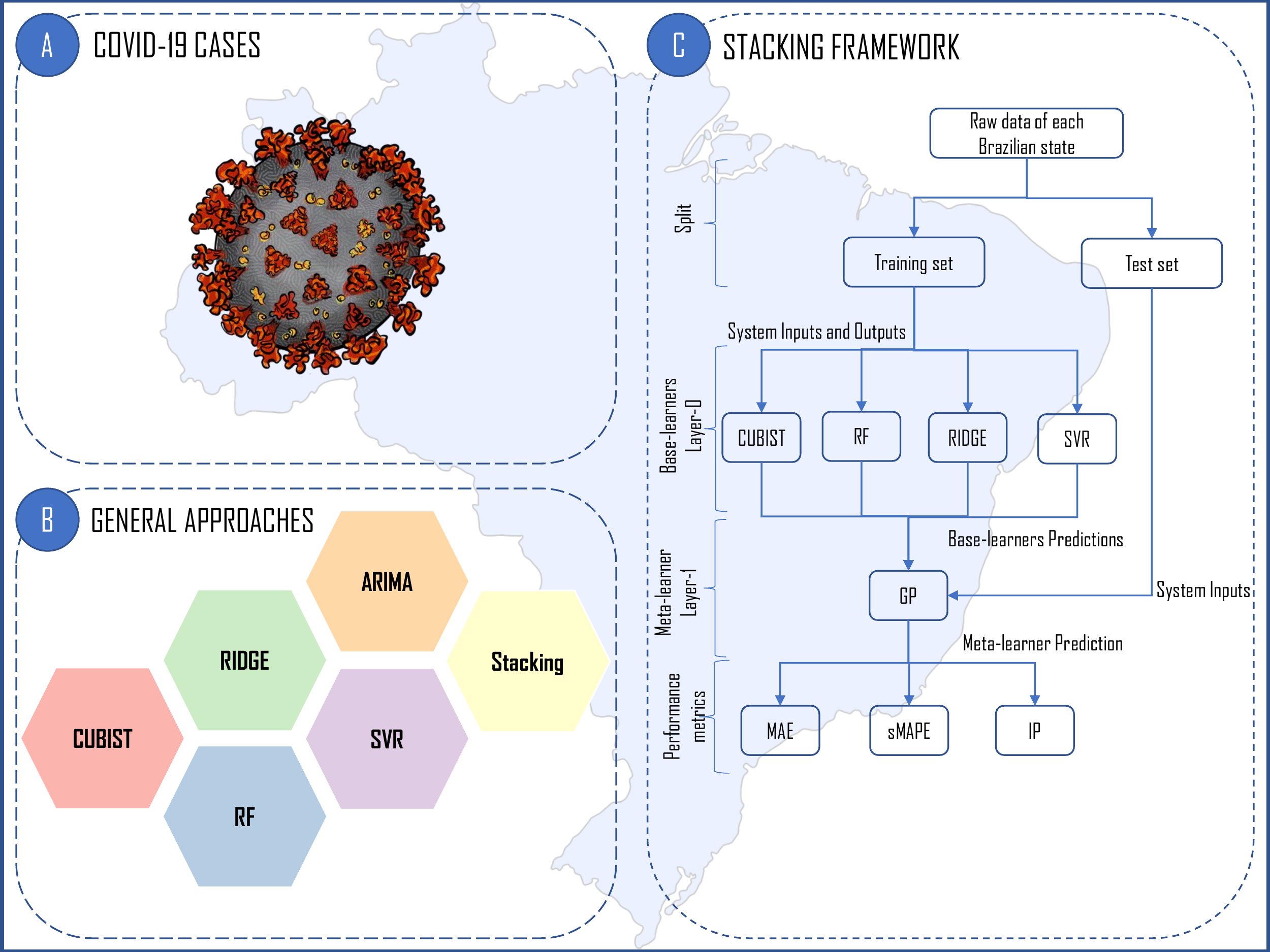}
    \caption{Proposed forecasting framework}
    \label{fig:flowchart}
\end{figure}

Figure \ref{fig:flowchart} presents the proposed forecasting framework.

    \section{Results \label{RES}}

This section describes the results of the developed experiments in forecasts out-of-sample (test set). First, Section \ref{sec:E1} compares the results of evaluated models over ten datasets and three forecasting horizons adopted. In Table \ref{tab:performancemeasure} in \ref{apendixA}, the best results regarding accuracy are presented in bold. Additionally, Figures \ref{fig:performance1} up to \ref{fig:performance3} illustrate the relation between observed and predicted values achieved by models with best set of performance measures depicted in Table \ref{tab:performancemeasure}, as well as box-plots for out-of-sample errors are illustrated in Figure \ref{fig:Boxplot}. 

\subsection{Performance Measures for compared models \label{sec:E1}}

In this section, the main results achieved by the best model regarding MAE and sMAPE criteria are presented for short-term forecasting multi-days- ahead of cumulative cases of COVID-19 from ten Brazilian states. 

\begin{itemize}
    \item AM: In this state, CUBIST, and RIDGE approaches could be considered to forecasting COVID-19 cases. In fact, in respect to ODA and TDA, CUBIST outperforms models, while for SDA the RIDGE achieves better accuracy regarding MAE and sMAPE than others. The improvement in the MAE for ODA and TDA achieved by CUBIST ranges between 6.58\% - 92.77\%, and 11.39\% - 88.54\%, respectively. Through sMAPE analysis, the RIDGE model outperforms other models, and this criterion is reduced in the range of 16.46\% - 91.88\%, for SDA horizon. 
    
    \item BA, MG, RS, and SP: For these states, in all forecasting windows, the SVR approach achieved better accuracy than other models, for both MAE and sMAPE criteria in the multi-days-ahead forecasting task of the confirmed number of COVID-19. In fact, the improvement in sMAPE is ranged in 13.26\% - 95.11\%, 4.23\% - 94.88\%, and 38.59\% - 95.24\%, respectively, in ODA, TDA, and SDA forecasting horizons. Moreover, the same behavior is observed when the improvement in sMAPE criterion is obtained. 
    
    \item CE and RN: In the CE state, the ARIMA model has a better performance in the forecasting out-of-sample than other models for ODA and TDA time windows. In this aspect, for MAE criterion, the improvement is ranged between 72.36\% - 98.03\%, and 45.93\% - 92.40\%, for ODA, and TDA time windows, respectively. For sMAPE, the improvement on ODA, and TDA horizons is 65.06\% - 97.84\%, and 32.81\% - 92.53\%, respectively. The SVR has better results than  ARIMA model for SDA. Considering the RN state, the same analysis is developed for ODA, and TDA horizons. The exception to the SDA horizon, in which the CUBIST model has better effectiveness in the MAE and sMAPE criteria than remain models.
    
    \item PR, RJ, and SC: For these states localized into the south region (PR and SC) and southeast region (RJ) of Brazil, the most appropriate approach to forecast cumulative cases of COVID-19 is the stacking ensemble, exception in ODA horizon, when ARIMA model has better results. Stacking overcomes the drawback of single models and achieves the best accuracy than other models. In fact, for these states, the improvement in MAE and sMAPE are between 14.01\% - 94.68\%, and 17.48\% - 95.41\%, respectively, for ODA horizon. The improvement in order forecasting horizons presents the same behavior of ODA, with the greatest magnitude of improvement for TDA and SDA. 
\end{itemize}

\textbf{Remark:} In this experiment, 180 scenarios (10 datasets, 3 forecasting horizons, and 6 models) were evaluated for the task of forecasting cumulative COVID-19 cases. In an overview, the best models for each state, obtained sMAPE ranged between 0.87\% - 3.51\%, 1.02\% - 5.63\%, and 0.95\% - 6.90\% for ODA, TDA, and SDA forecasting, respectively. The ranking of models in all scenarios is SVR, stacking ensemble, ARIMA, CUBIST, RIDGE, and RF models. In contrast to finds of \cite{BENVENUTO2020105340}, for the datasets evaluated in this paper, ARIMA modelling was effective in some situation for very-short horizons When the horizon is SDA, ARIMA model has worst performance than most of compared models. However, for ODA the applications are limited. From a broader perspective, the efficiency of SVR is due to its ability to deal with small size dataset, while the stacking ensemble combines the advantages of several single models to learn the data behavior and obtain forecasts similar to observed values. On the other hand, the difficulty of the RF model to forecasting cumulative COVID-19 cases could be attributed to the fact that this approach requires more observations to effectively learn the data pattern. 

\begin{figure}[htb!]
    \centering
    \subfloat[AM \label{fig:pred_AM}]{
    \includegraphics[width=0.45\linewidth]{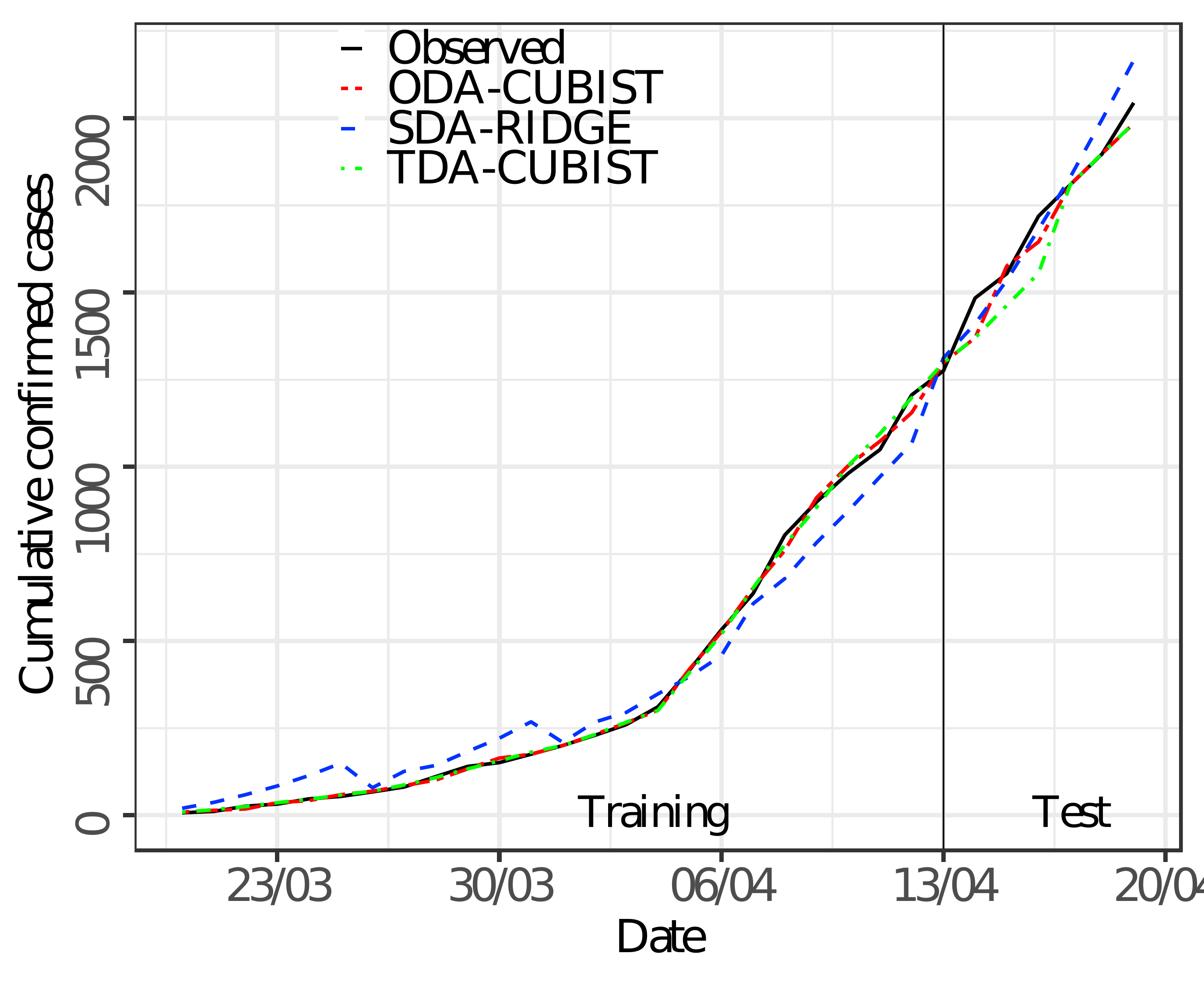}}
    \subfloat[BA\label{fig:pred_BA}]{
    \includegraphics[width=0.45\linewidth]{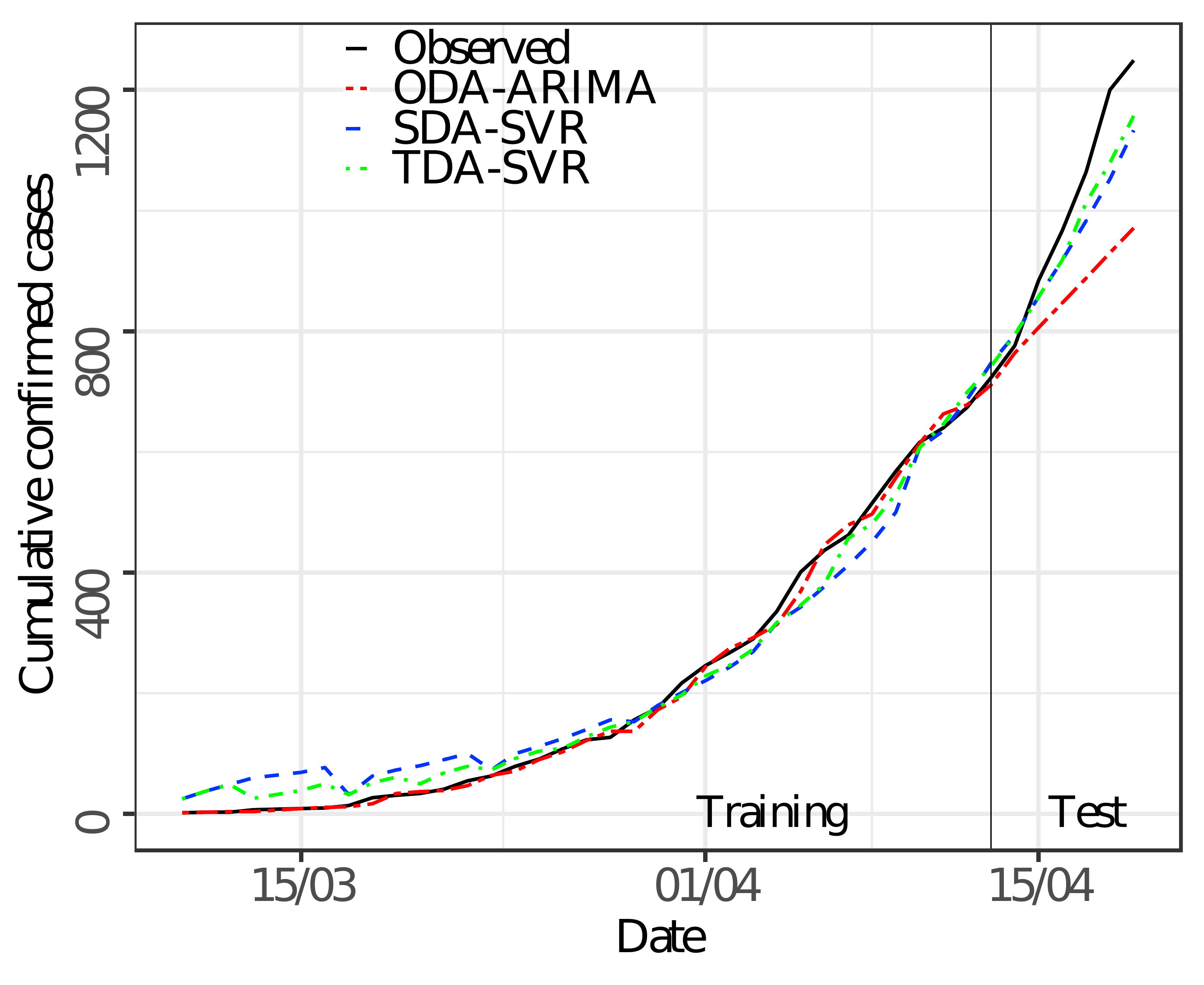}}\\
    

\centering
    \subfloat[CE \label{fig:pred_CE}]{
    \includegraphics[width=0.45\linewidth]{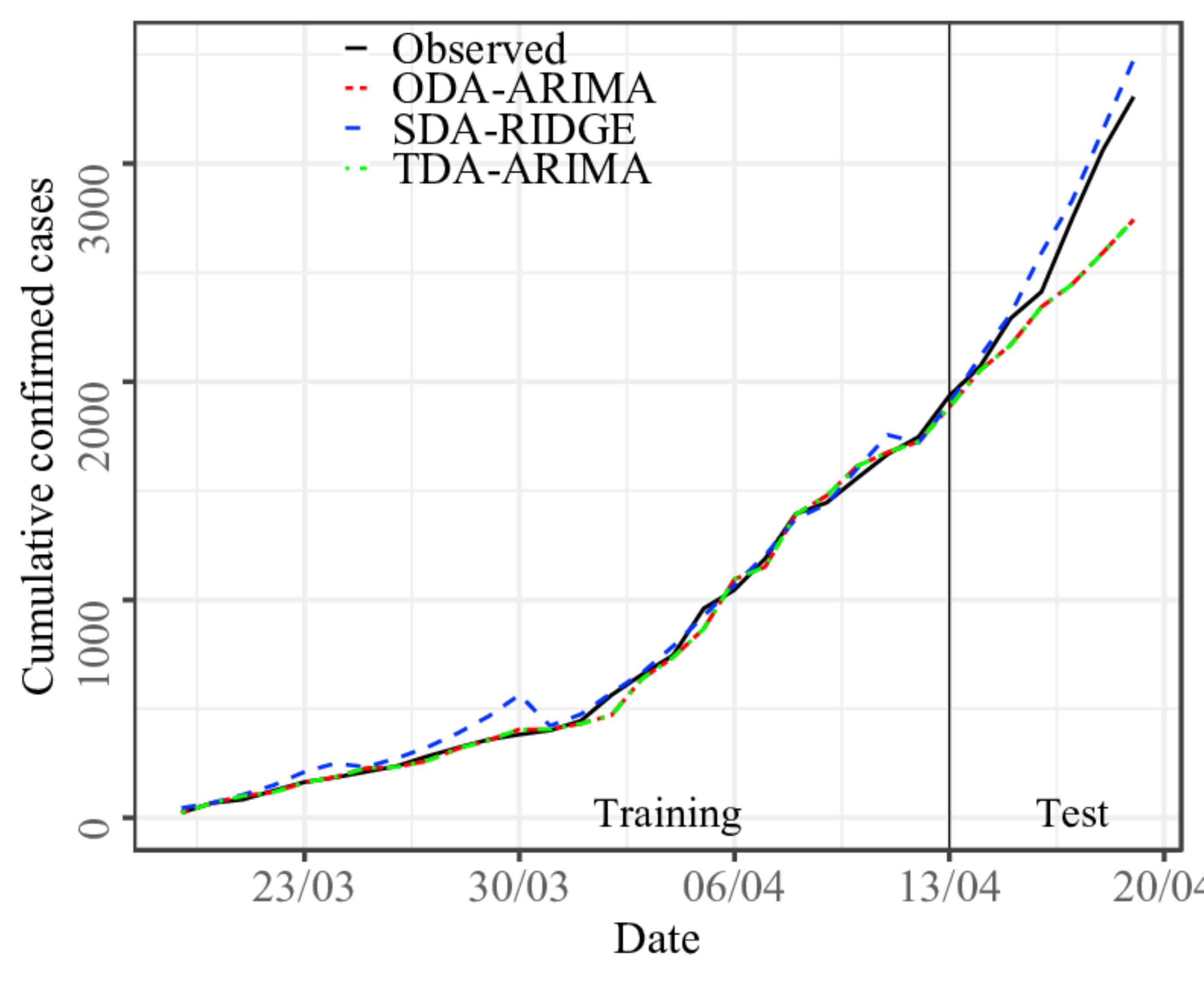}}
    \subfloat[MG\label{fig:pred_MG}]{
    \includegraphics[width=0.45\linewidth]{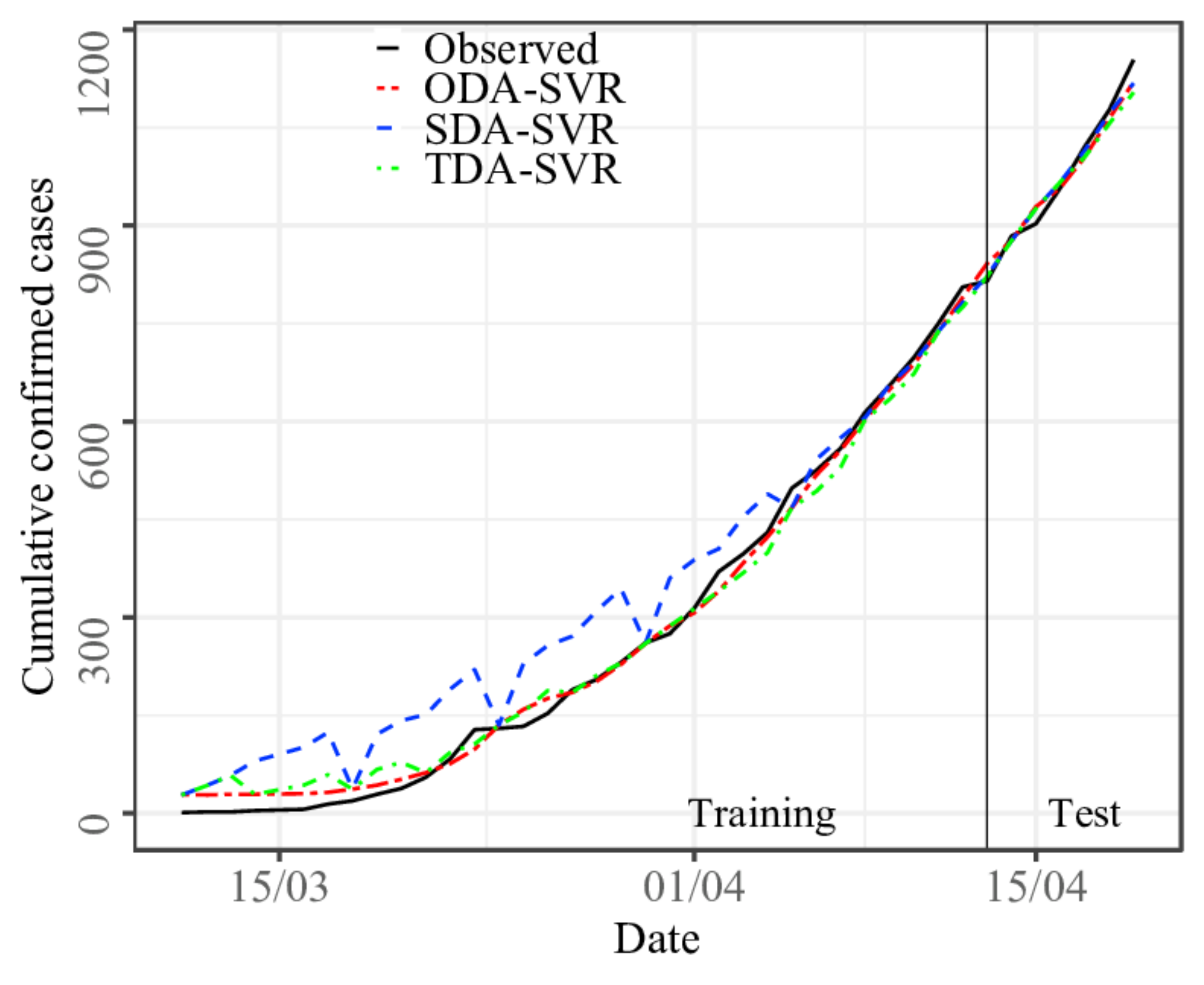}}
     \caption{Predicted versus observed cumulative confirmed cases of COVID-19 for AM, BA, CE, and MG states}
     \label{fig:performance1}
 \end{figure}   
 
 \begin{figure}[htb!]
    \centering
    \subfloat[PR \label{fig:pred_PR}]{
    \includegraphics[width=0.42\linewidth]{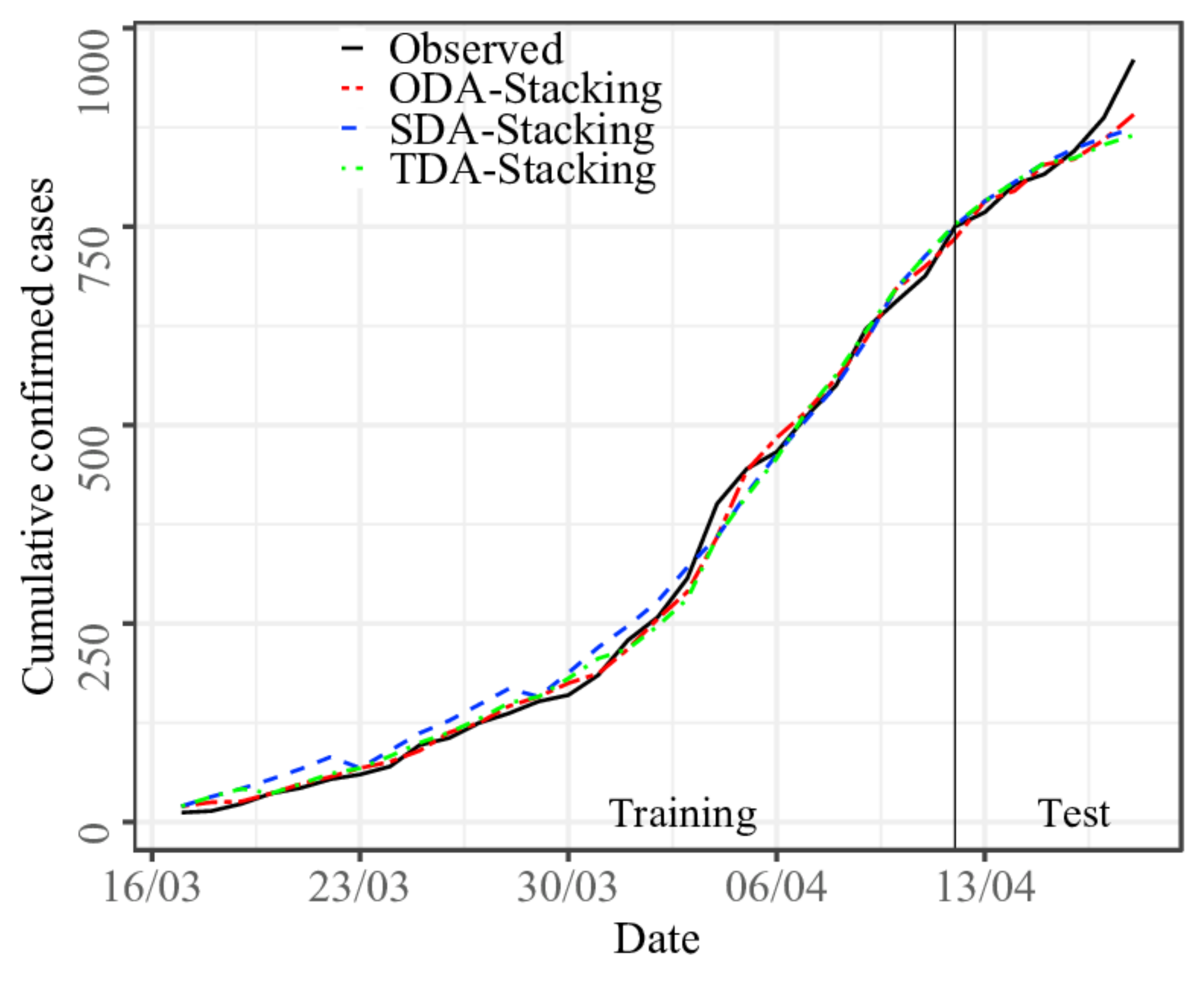}}
    \subfloat[RJ\label{fig:pred_RJ}]{
    \includegraphics[width=0.42\linewidth]{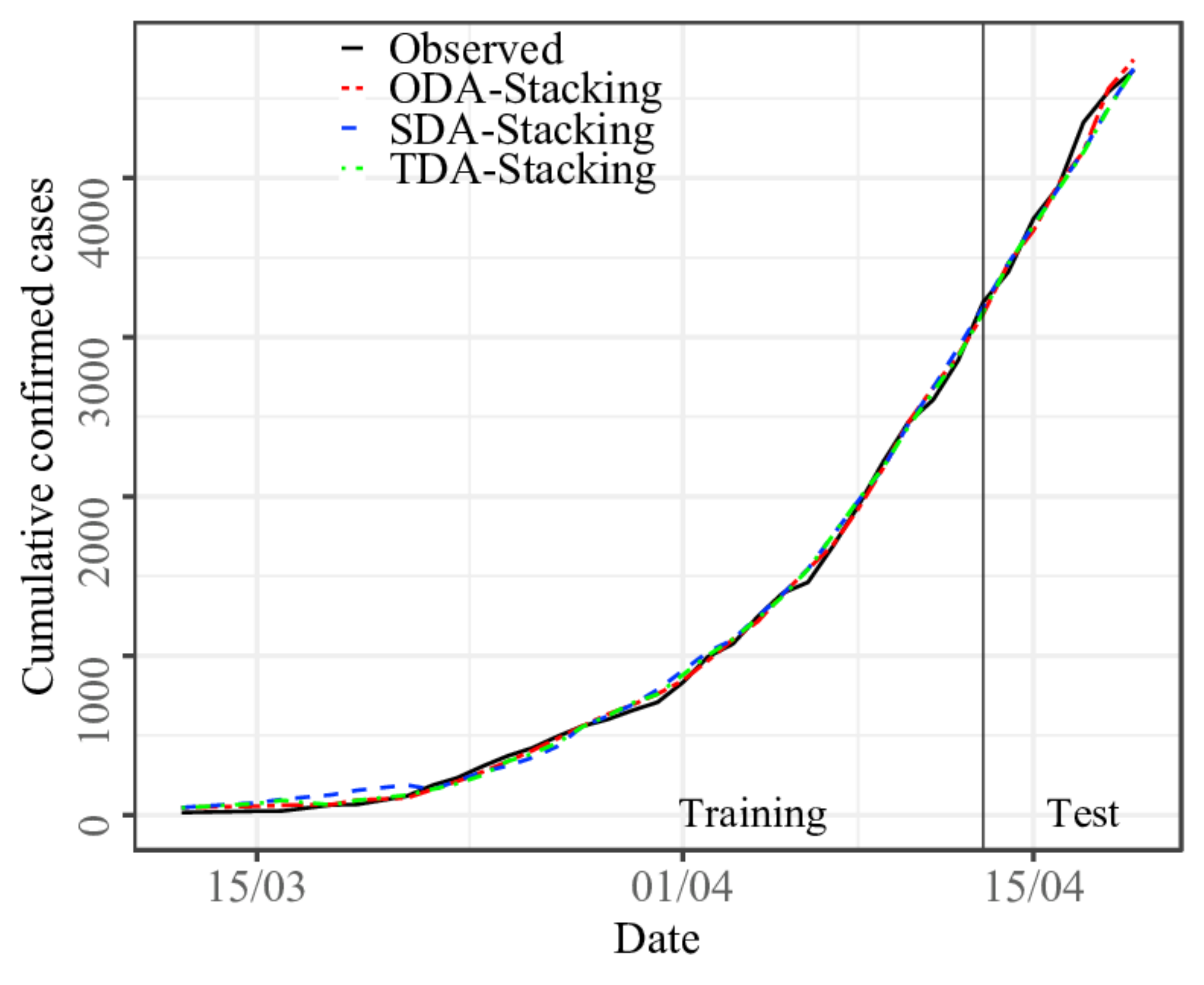}}\\
    
    \subfloat[RN \label{fig:pred_RN}]{
    \includegraphics[width=0.42\linewidth]{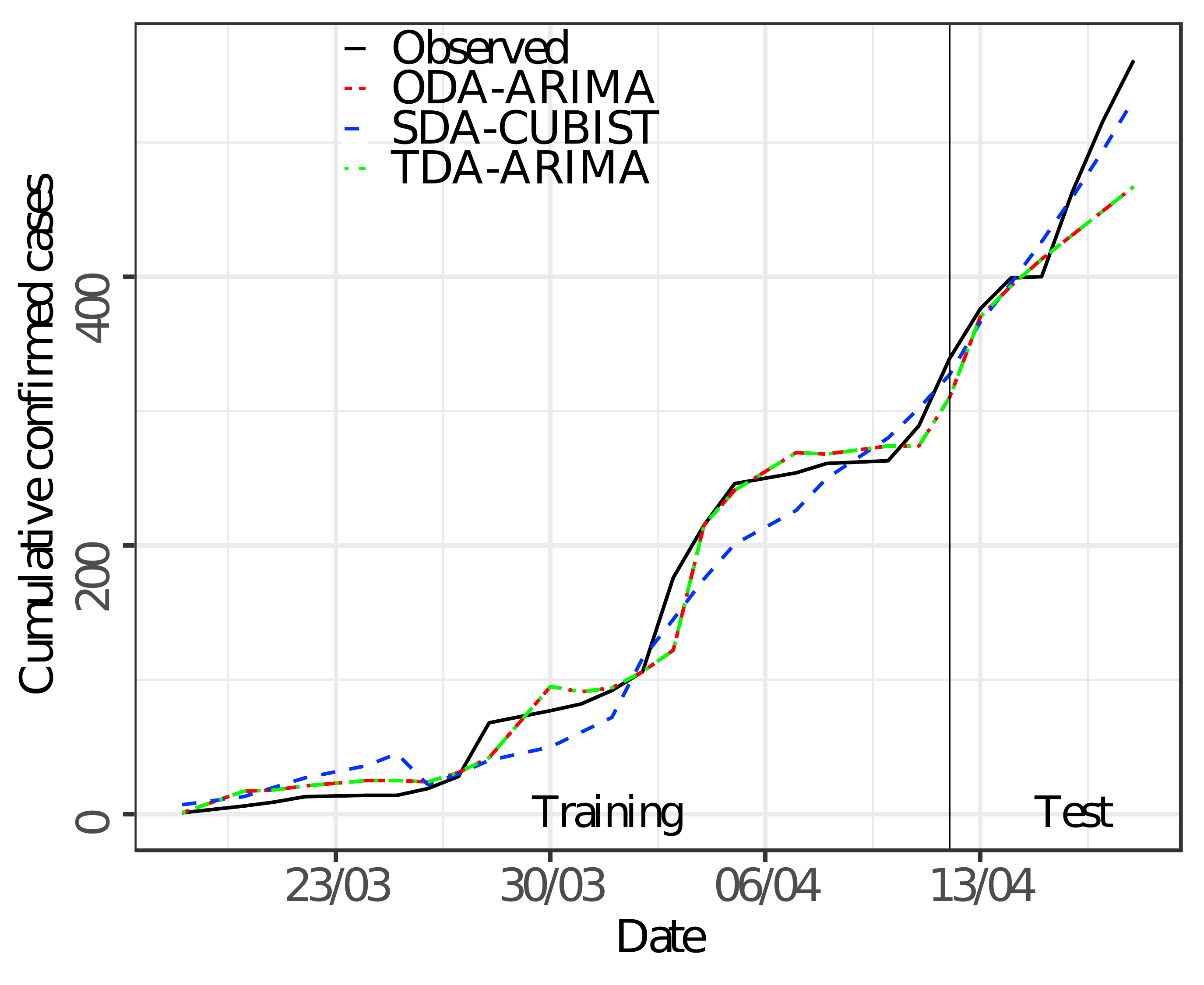}}
    \subfloat[RS\label{fig:pred_RS}]{
    \includegraphics[width=0.42\linewidth]{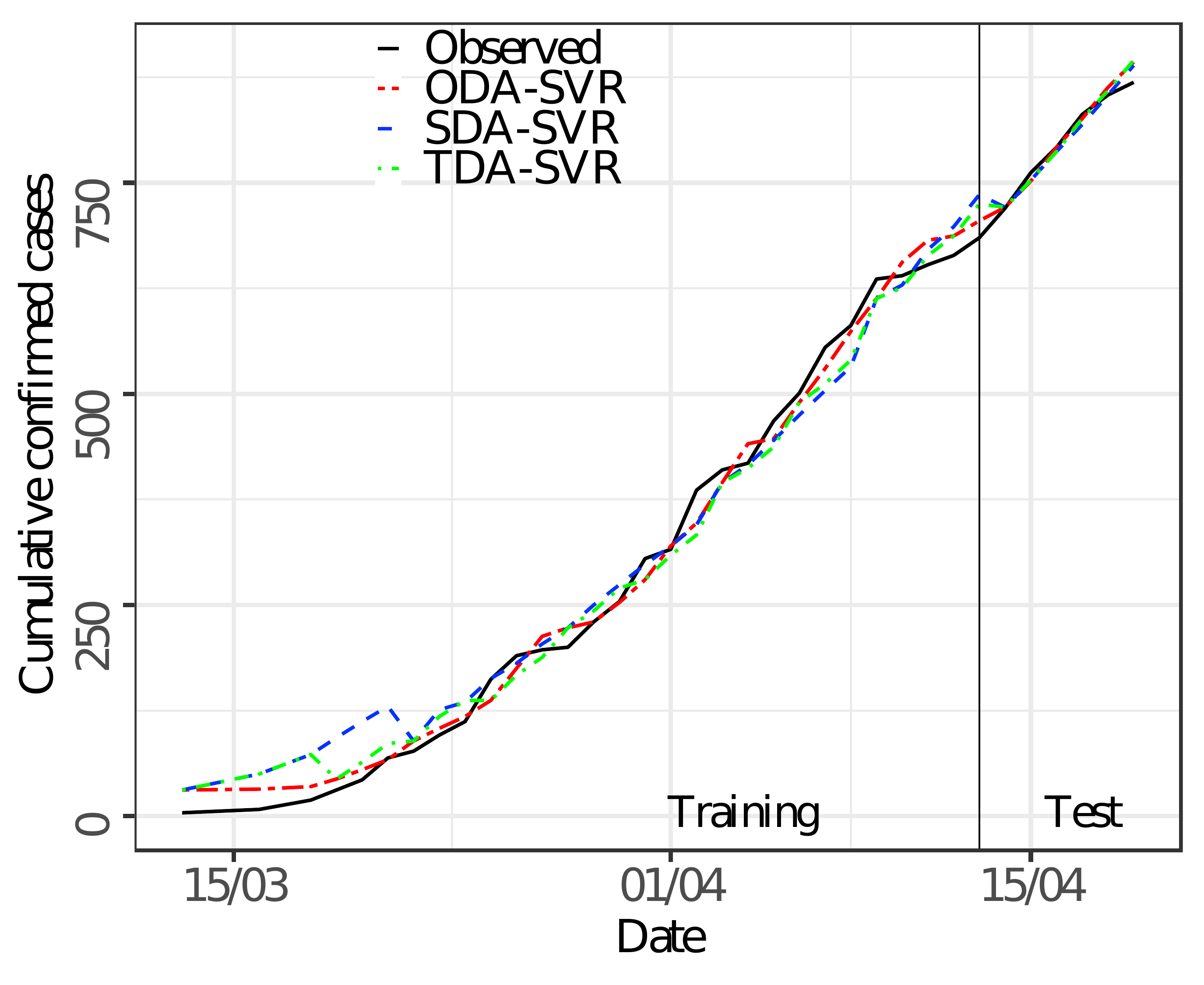}} \\
    

    \subfloat[SC \label{fig:pred_SC}]{
    \includegraphics[width=0.42\linewidth]{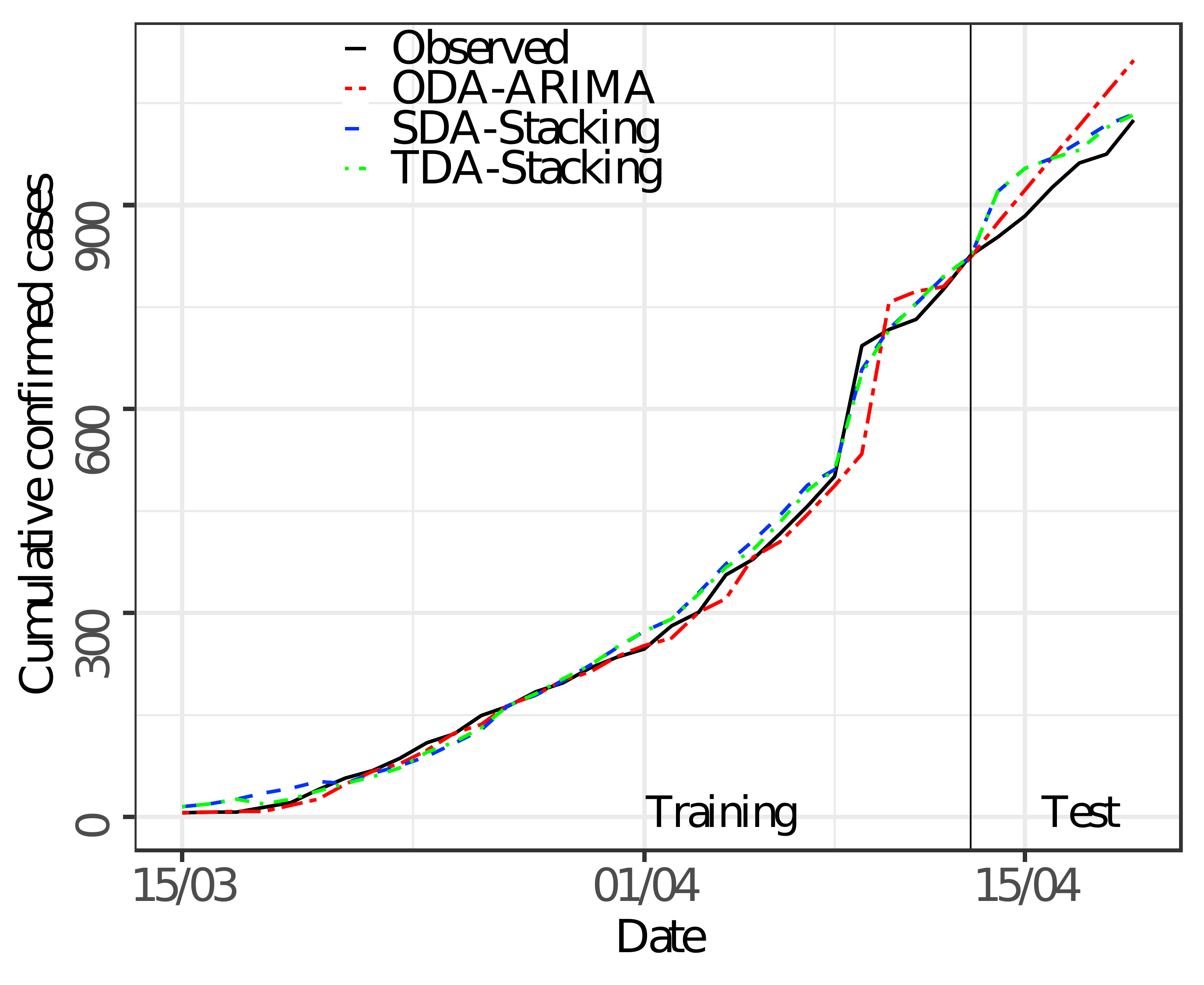}}
    \subfloat[SP\label{fig:pred_SP}]{
    \includegraphics[width=0.42\linewidth]{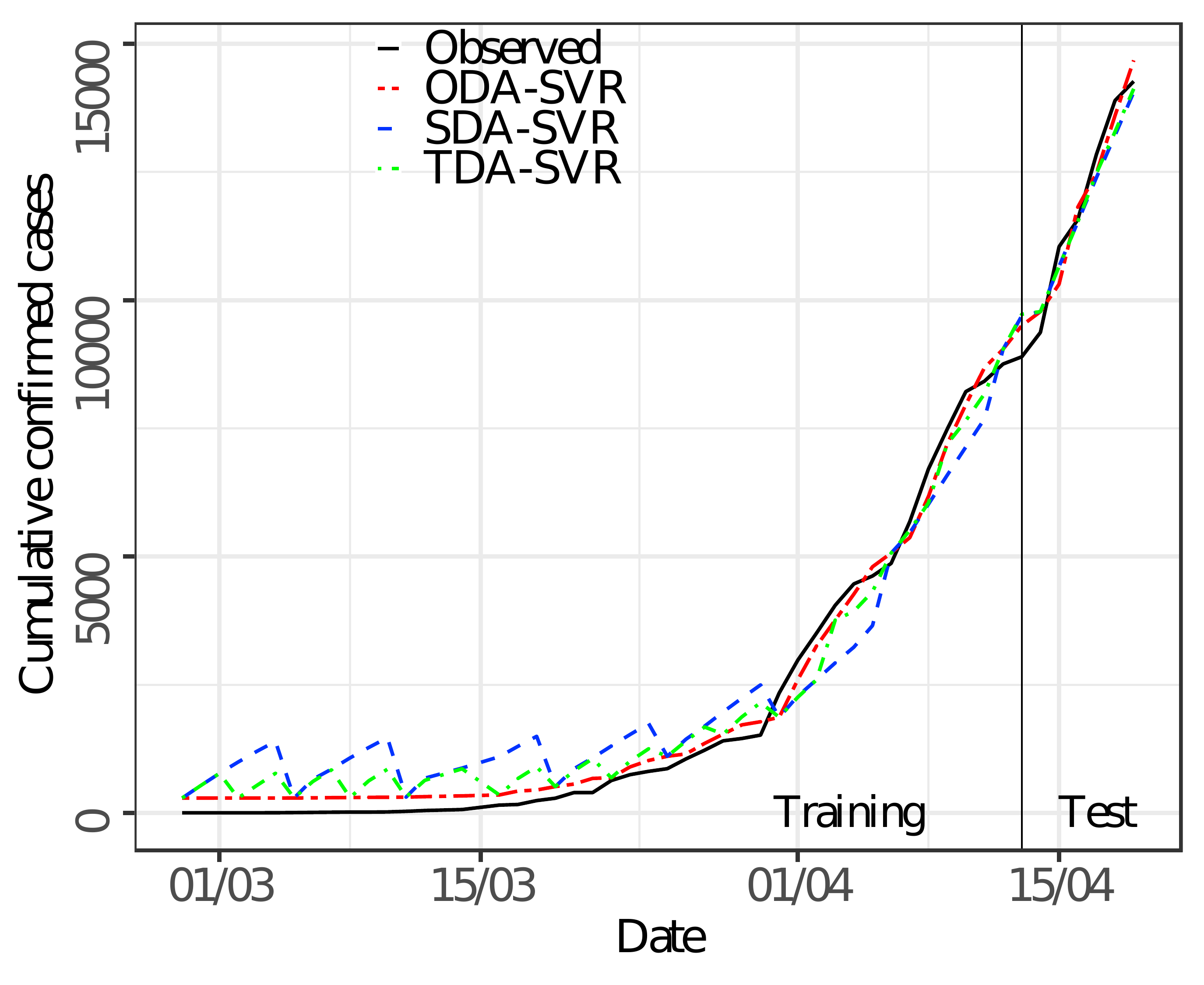}}

    \caption{Predicted versus observed cumulative confirmed cases of COVID-19 for PR, RJ, RN, RS, SC, and SP states}
    \label{fig:performance3}
\end{figure}

According to the information depicted in Figures \ref{fig:performance1} and \ref{fig:performance3} it is possible to identify that the behavior of the data is learned by the evaluated models, which can forecasting compatible cases with the observed values. The good performance obtained in the training phase persists in the test stage. In the Figures \ref{fig:pred_AM} and \ref{fig:pred_RN} the models, RIDGE and CUBIST, as well as in Figures \ref{fig:pred_MG} and \ref{fig:pred_SP} SVR presented difficulties to capture the variability of the first observations. The dataset is reduced for all states, which justifies the difficulties of the mathematical models to learn the behavior.

\begin{figure}[htb!]
    \centering
    \includegraphics[width=.8\linewidth,angle=90]{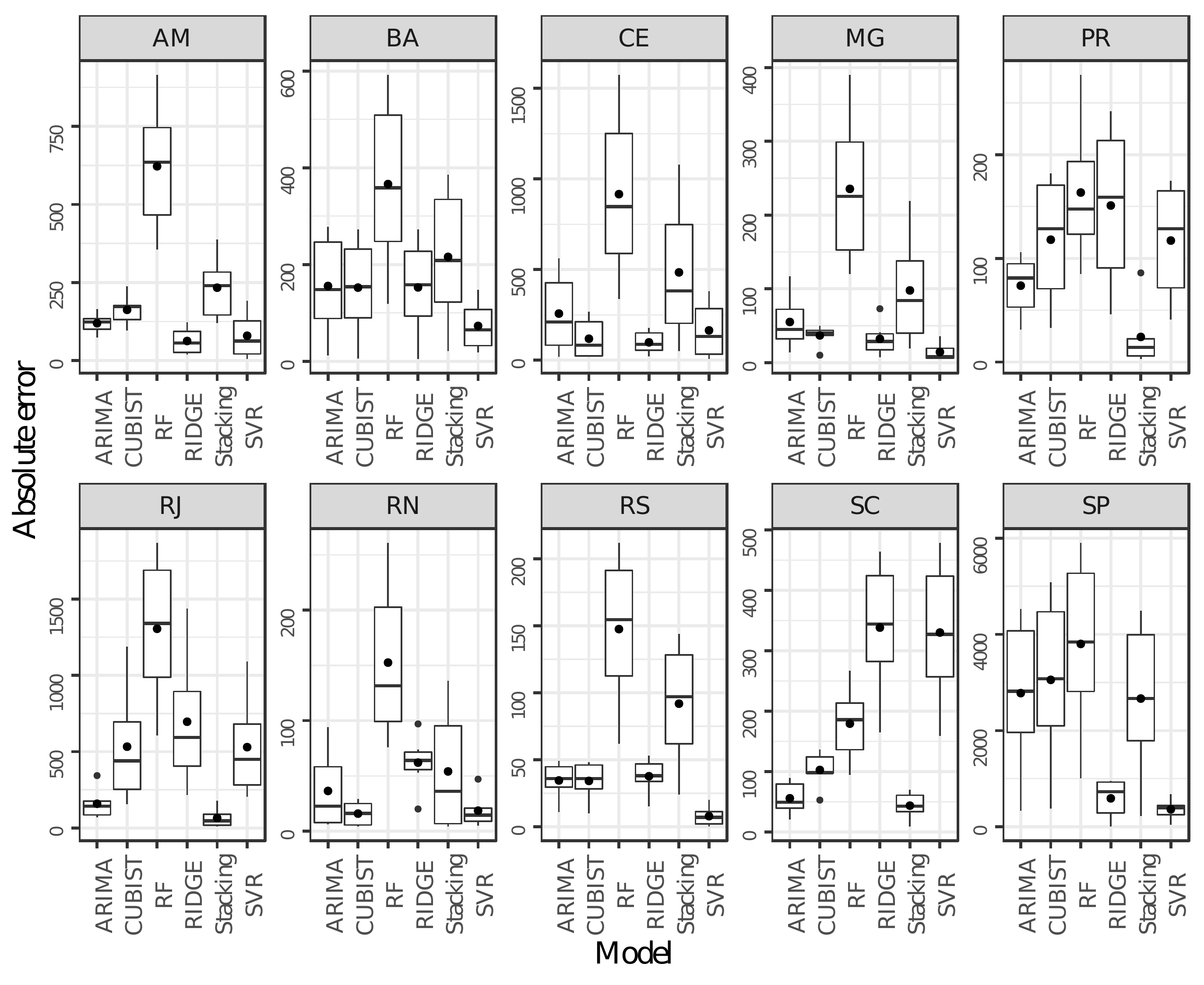}
    \caption{Box-plot for absolute error according to model and state for COVID-19 forecasting up to SDA}
    \label{fig:Boxplot}
\end{figure}

Figure \ref{fig:Boxplot} shows the box-plots of out-of-sample forecasting errors in the SDA horizon for each model and dataset used. This horizon is chosen to analysis due to the recursive strategy adopted, once the errors increase according to the growth of the forecasting horizon. The box diagram depicts the variation of absolute errors for each model, which reflects the stability of each model. In this context, the dots out of boxes are considered outliers errors, and the black dot inside of the box is the MAE for each model.

Through the box-plot analysis, boxes with lower size indicate models with lower variation in the errors, and the results presented in Table \ref{tab:performancemeasure} are corroborated by the depicted in Figure \ref{fig:Boxplot}. Models with lower errors also reach better stability, which means that the most suitable modelling for each state can maintain a learning pattern, achieving homogeneous prediction errors.

    \section{Conclusion and Future Research \label{CONC}}

In this paper, six machine learning approaches named CUBIST, RF, RIDGE, SVR, and stacking ensemble, as well as ARIMA statistical model, were employed in the task of forecasting one, three, and six-days-ahead the COVID-19 cumulative confirmed cases in ten Brazilian states with a high daily incidence. The COVID-19 cumulative confirmed cases for AM, BA, CE, MG, PR, RJ, RN, RS, SC, and SP states were used. The IP, MAE, and sMAPE criteria were adopted to evaluate the performance of the compared approaches. Moreover, the stability of out-of-sample errors was evaluated through box-plots.

In respect of obtained results, it is possible to infer that SVR and stacking-ensemble learning model are suitable tools to forecast COVID-19 cases for most of the adopted states, once that these approaches were able to learn the nonlinearities inherent to the evaluated epidemiological time series. Also, ARIMA can be considered in some aspects for ODA, while CUBIST and RIDGE models deserve attention for the development of this task in TDA and SDA time windows. Therefore, the ranking of models in all scenarios is SVR, stacking ensemble, ARIMA, CUBIST, RIDGE, and RF models. However, even though the models discussed in this paper presented forecasting cases similar to those observed, they should be used cautiously. This fact is attributed to the chaotic dynamics of the analyzed data, as well as the diversity of exogenous factors that can affect the daily notifications of COVID-19.

For future works, it is intended (i) to adopt deep learning approaches combined to stacking ensemble, (ii) to employ copulas functions for data augmentation dealing with small samples, (iii) to use multi-objective optimization to tune hyperparameters of adopted models, (iv) to adopt set of features which can help to explain the future cases of the COVID-19.

    \section*{CRediT Author Statement}

\textbf{Matheus Henrique Dal Molin Ribeiro:} Conceptualization, Methodology, Formal analysis, Validation, Writing - Original Draft, Writing - Review \& Editing.
\textbf{Ramon Gomes da Silva:} Conceptualization, Methodology, Formal analysis, Validation, Writing - Original Draft, Writing - Review \& Editing.
\textbf{Viviana Cocco Mariani:} Conceptualization, Writing - Review \& Editing.
\textbf{Leandro dos Santos Coelho:} Conceptualization, Writing - Review \& Editing.
    \section*{Declaration of Competing Interest}

The authors declare that they have no known competing financial interests or personal relationships that could have appeared to influence the work reported in this paper.
    \section*{Acknowledgments}

The authors would like to thank the National Council of Scientific and Technologic Development of Brazil -- CNPq (Grants number: 307958/2019-1-PQ, 307966/2019-4-PQ, 404659/2016-0-Univ, 405101/2016-3-Univ),  PRONEX `\textit{Funda\c{c}\~ao Arauc\'aria}' 042/2018, and \textit{Coordena\c{c}\~ao de Aperfei\c{c}oamento de Pessoal de N\'ivel Superior - Brasil} (CAPES) - Finance Code 001 for financial support of this work.
    
    \bibliographystyle{Misc/model3-num-names.bst}
    \bibliography{Misc/theBiblio}
    
    \appendix
    \newpage
    \section{Performance Measures \label{apendixA}}

Table \ref{tab:performancemeasure} presents the performance measures for each model in each state and forecasting horizon.

\setcounter{table}{0}

\begin{table}[htb!]
\tiny
\centering
\caption{Performance measures for each evaluated model}
\label{tab:performancemeasure}
\resizebox{.8\textwidth}{!}{%
\begin{tabular}{ccl|cccccc}
\hline
\multirow{2}{*}{State} & \multicolumn{1}{c}{\multirow{2}{*}{\begin{tabular}[c]{@{}c@{}}Forecasting\\ Horizon\end{tabular}}} & \multirow{2}{*}{Criteria} & \multicolumn{6}{c}{Model}                                                                                                                                            \\ \cline{4-9} 
                       & \multicolumn{1}{c}{}                                                                               &                           & \multicolumn{1}{c}{ARIMA} & \multicolumn{1}{c}{CUBIST} & \multicolumn{1}{c}{RF} & \multicolumn{1}{c}{RIDGE} & \multicolumn{1}{c}{Stacking} & \multicolumn{1}{c}{SVR} \\ \hline
\multirow{6}{*}{AM}    & \multirow{2}{*}{ODA}                                                                    & MAE                       & 95                        & \textbf{45}                & 622.17                 & 48.17                     & 121.5                        & 56.33                   \\
                       &                                                                                                    & sMAPE                     & 6.61\%                    & \textbf{2.80\%}            & 42.50\%                & 2.83\%                    & 7.13\%                       & 3.18\%                  \\
                       & \multirow{2}{*}{TDA}                                                                 & MAE                       & 101.33                    & \textbf{71.33}             & 622.17                 & 83.67                     & 176.67                       & 80.5                    \\
                       &                                                                                                    & sMAPE                     & 6.55\%                    & 4.50\%                     & 42.50\%                & 4.49\%                    & 10.47\%                      & \textbf{4.19\%}         \\
                       & \multirow{2}{*}{SDA}                                                                   & MAE                       & 119.17                    & 162.17                     & 622.17                 & \textbf{62.33}            & 233.17                       & 79.17                   \\
                       &                                                                                                    & sMAPE                     & 6.97\%                    & 9.55\%                     & 42.50\%                & \textbf{3.45\%}           & 13.87\%                      & 4.13\%                  \\ \hline
\multirow{6}{*}{BA}    & \multirow{2}{*}{ODA}                                                                    & MAE                       & \textbf{12}               & 93.83                      & 366.33                 & 45.33                     & 107.67                       & 42.33                   \\
                       &                                                                                                    & sMAPE                     & \textbf{1.56\%}           & 9.16\%                     & 42.02\%                & 4.36\%                    & 10.68\%                      & 4.15\%                  \\
                       & \multirow{2}{*}{TDA}                                                                 & MAE                       & 70                        & 132                        & 366.33                 & 74.33                     & 171.67                       & \textbf{59.67}          \\
                       &                                                                                                    & sMAPE                     & 8.00\%                    & 12.92\%                    & 42.02\%                & 7.46\%                    & 17.32\%                      & \textbf{5.63\%}         \\
                       & \multirow{2}{*}{SDA}                                                                   & MAE                       & 155.67                    & 152.33                     & 366.33                 & 152.83                    & 215.83                       & \textbf{73.17}          \\
                       &                                                                                                    & sMAPE                     & 15.41\%                   & 15.08\%                    & 42.02\%                & 15.16\%                   & 22.25\%                      & \textbf{6.90\%}         \\ \hline
\multirow{6}{*}{CE}    & \multirow{2}{*}{ODA}                                                                    & MAE                       & \textbf{18}               & 65.17                      & 916                    & 70.33                     & 220.83                       & 87.67                   \\
                       &                                                                                                    & sMAPE                     & \textbf{0.87\%}           & 2.49\%                     & 40.28\%                & 2.81\%                    & 8.20\%                       & 3.17\%                  \\
                       & \multirow{2}{*}{TDA}                                                                 & MAE                       & \textbf{69.66}            & 128.83                     & 916                    & 149.83                    & 382.17                       & 136.67                  \\
                       &                                                                                                    & sMAPE                     & \textbf{3.01\%}           & 4.48\%                     & 40.28\%                & 5.39\%                    & 14.48\%                      & 4.78\%                  \\
                       & \multirow{2}{*}{SDA}                                                                   & MAE                       & 257                       & 118.17                     & 916                    & \textbf{98.17}                     & 484.33                       & 164.17         \\
                       &                                                                                                    & sMAPE                     & 9.34\%                    & 4.11\%                     & 40.28\%                & \textbf{3.52\%}                    & 18.78\%                      & 5.77\%         \\ \hline
\multirow{6}{*}{MG}    & \multirow{2}{*}{ODA}                                                                    & MAE                       & 32                        & 17.5                       & 235.5                  & 24.33                     & 56.5                         & \textbf{16}             \\
                       &                                                                                                    & sMAPE                     & 3.63\%                    & 1.81\%                     & 26.21\%                & 2.50\%                    & 5.59\%                       & \textbf{1.57\%}         \\
                       & \multirow{2}{*}{TDA}                                                                 & MAE                       & 26                        & 21.33                      & 235.5                  & 21.67                     & 78.17                        & \textbf{21}             \\
                       &                                                                                                    & sMAPE                     & 3.08\%                    & 2.20\%                     & 26.21\%                & 2.13\%                    & 7.81\%                       & \textbf{2.04\%}         \\
                       & \multirow{2}{*}{SDA}                                                                   & MAE                       & 55                        & 36.83                      & 235.5                  & 32.17                     & 97.83                        & \textbf{14.33}          \\
                       &                                                                                                    & sMAPE                     & 5.43\%                    & 3.58\%                     & 26.21\%                & 3.14\%                    & 9.88\%                       & \textbf{1.41\%}         \\ \hline
\multirow{6}{*}{PR}    & \multirow{2}{*}{ODA}                                                                    & MAE                       & 31                        & 27.33                      & 163.5                  & 38                        & \textbf{23.5}                & 35.33                   \\
                       &                                                                                                    & sMAPE                     & 3.96\%                    & 3.26\%                     & 21.09\%                & 4.50\%                    & \textbf{2.69\%}              & 4.18\%                  \\
                       & \multirow{2}{*}{TDA}                                                                 & MAE                       & 51.66                     & 57.33                      & 163.5                  & 76.5                      & \textbf{28.17}               & 60.17                   \\
                       &                                                                                                    & sMAPE                     & 6.21\%                    & 6.56\%                     & 21.09\%                & 8.61\%                    & \textbf{3.21\%}              & 6.89\%                  \\
                       & \multirow{2}{*}{SDA}                                                                   & MAE                       & 73.67                     & 118                        & 163.5                  & 151                       & \textbf{24.17}               & 117.17                  \\
                       &                                                                                                    & sMAPE                     & 8.20\%                    & 12.56\%                    & 21.09\%                & 15.75\%                   & \textbf{2.75\%}              & 12.53\%                 \\ \hline
\multirow{6}{*}{RJ}    & \multirow{2}{*}{ODA}                                                                    & MAE                       & 110                       & 165.5                      & 1305.67                & 273.67                    & \textbf{69.5}                & 360.83                  \\
                       &                                                                                                    & sMAPE                     & 3.17\%                    & 3.82\%                     & 37.06\%                & 6.25\%                    & \textbf{1.70\%}              & 8.09\%                  \\
                       & \multirow{2}{*}{TDA}                                                                 & MAE                       & 120                       & 275.67                     & 1305.67                & 462.83                    & \textbf{68}                  & 429.33                  \\
                       &                                                                                                    & sMAPE                     & 3.18\%                    & 6.24\%                     & 37.06\%                & 10.20\%                   & \textbf{1.65\%}              & 9.49\%                  \\
                       & \multirow{2}{*}{SDA}                                                                   & MAE                       & 158.33                    & 532.67                     & 1305.67                & 696.17                    & \textbf{65.17}               & 529.5                   \\
                       &                                                                                                    & sMAPE                     & 3.67\%                    & 11.34\%                    & 37.06\%                & 14.67\%                   & \textbf{1.58\%}              & 11.43\%                 \\ \hline
\multirow{6}{*}{RN}    & \multirow{2}{*}{ODA}                                                                    & MAE                       & \textbf{6}                & 17                         & 152.5                  & 24.83                     & 30.33                        & 18.33                   \\
                       &                                                                                                    & sMAPE                     & \textbf{1.61\%}           & 3.87\%                     & 39.28\%                & 5.56\%                    & 6.45\%                       & 4.14\%                  \\
                       & \multirow{2}{*}{TDA}                                                                 & MAE                       & \textbf{8.33}             & 30.83                      & 152.5                  & 37.67                     & 54                           & 35.5                    \\
                       &                                                                                                    & sMAPE                     & \textbf{2.11\%}           & 6.54\%                     & 39.28\%                & 8.51\%                    & 11.66\%                      & 7.69\%                  \\
                       & \multirow{2}{*}{SDA}                                                                   & MAE                       & 36.33                     & \textbf{15.83}                      & 152.5                  & 62                        & 54                           & 18.5           \\
                       &                                                                                                    & sMAPE                     & 7.61\%                    & \textbf{3.42\%}\%                     & 39.28\%                & 12.76\%                   & 11.66\%                      & 4.15\%         \\ \hline
\multirow{6}{*}{RS}    & \multirow{2}{*}{ODA}                                                                    & MAE                       & 12                        & 12.83                      & 146.67                 & 11.33                     & 45.5                         & \textbf{8.17}           \\
                       &                                                                                                    & sMAPE                     & 1.64\%                    & 1.62\%                     & 19.82\%                & 1.43\%                    & 5.76\%                       & \textbf{0.97\%}         \\
                       & \multirow{2}{*}{TDA}                                                                 & MAE                       & 24                        & 19.17                      & 147.33                 & 18.67                     & 71.33                        & \textbf{8.5}            \\
                       &                                                                                                    & sMAPE                     & 3.22\%                    & 2.47\%                     & 19.92\%                & 2.42\%                    & 9.14\%                       & \textbf{1.02\%}         \\
                       & \multirow{2}{*}{SDA}                                                                   & MAE                       & 34.5                      & 34.17                      & 147.5                  & 37.67                     & 91.83                        & \textbf{7.83}           \\
                       &                                                                                                    & sMAPE                     & 4.31\%                    & 4.26\%                     & 19.95\%                & 4.74\%                    & 11.89\%                      & \textbf{0.95\%}         \\ \hline
\multirow{6}{*}{SC}    & \multirow{2}{*}{ODA}                                                                    & MAE                       & \textbf{21}               & 93.67                      & 179.5                  & 180.5                     & 33.83                        & 177.67                  \\
                       &                                                                                                    & sMAPE                     & \textbf{2.43\%}           & 9.66\%                     & 20.97\%                & 17.53\%                   & 3.66\%                       & 17.27\%                 \\
                       & \multirow{2}{*}{TDA}                                                                 & MAE                       & 44.33                     & 100.33                     & 179.5                  & 277                       & \textbf{41}                  & 257.33                  \\
                       &                                                                                                    & sMAPE                     & 4.76\%                    & 10.30\%                    & 20.97\%                & 25.34\%                   & \textbf{4.39\%}              & 23.79\%                 \\
                       & \multirow{2}{*}{SDA}                                                                   & MAE                       & 56                        & 102.83                     & 179.5                  & 338.5                     & \textbf{43.83}               & 330.33                  \\
                       &                                                                                                    & sMAPE                     & 5.65\%                    & 10.53\%                    & 20.97\%                & 29.95\%                   & \textbf{4.68\%}              & 29.23\%                 \\ \hline
\multirow{6}{*}{SP}    & \multirow{2}{*}{ODA}                                                                    & MAE                       & 436                       & 1587                       & 3799                   & 537.33                    & 1363.83                      & \textbf{409}            \\
                       &                                                                                                    & sMAPE                     & 4.65\%                    & 13.47\%                    & 35.85\%                & 4.44\%                    & 11.44\%                      & \textbf{3.51\%}         \\
                       & \multirow{2}{*}{TDA}                                                                 & MAE                       & 1485.66                   & 2471.83                    & 3801                   & 579.17                    & 2243                         & \textbf{326.67}         \\
                       &                                                                                                    & sMAPE                     & 14.56\%                   & 21.81\%                    & 35.88\%                & 4.79\%                    & 19.47\%                      & \textbf{2.77\%}         \\
                       & \multirow{2}{*}{SDA}                                                                   & MAE                       & 2779                      & 3054.67                    & 3801.5                 & 591.83                    & 2665.83                      & \textbf{362.83}         \\
                       &                                                                                                    & sMAPE                     & 24.74\%                   & 27.60\%                    & 35.88\%                & 4.95\%                    & 23.55\%                      & \textbf{3.04\%}         \\ \hline
\end{tabular}
}
\end{table}
    \section{Hyperparameters \label{apendixB}}

Table \ref{tab:hyper}  presents the hyperparameters obtained by grid-search for the models employed in this paper. In the stacking modeling, for the GP meta-learner there is no hyperaparameter to be tuned.

\setcounter{table}{0}
\begin{table}[htb!]
\centering
    \caption{Hyperparameters selected by grid-search for each evaluated model}
 \label{tab:hyper}
\resizebox{\textwidth}{!}{%
\begin{tabular}{ccccccc}
\hline
\multirow{3}{*}{State} & \multicolumn{6}{c}{Model}                                                                                                                        \\ \cline{2-7} 
                       & ARIMA   & \multicolumn{2}{c}{CUBIST} & SVR  & RIDGE          & RF                                                                                \\ \cline{2-7} 
                       & \textit{(p,d,q)} & Committees   & Neighbors   & Cost & Regularization & \begin{tabular}[l]{@{}c@{}}Number of randomly \\ selected predictors\end{tabular} \\ \hline
AM                     & (1,2,0) & 10           & 5           & 1    & 3.16E-03       & 2                                                                                 \\
BA                     & (0,2,1) & 20           & 9           & 1    & 1E-04       & 2                                                                                 \\
CE                     & (2,2,1) & 1            & 9           & 1    & 0       & 4                                                                                 \\
MG                     & (0,2,1) & 1            & 9           & 1    & 1E-04       & 2                                                                                 \\
PR                     & (0,2,1) & 20           & 5           & 1    & 3.16E-03       & 3                                                                                 \\
RJ                     & (0,2,1) & 1            & 9           & 1    & 1E-04       & 3                                                                                 \\
RN                     & (1,1,0) & 1            & 9           & 1    & 3.16E-03       & 5                                                                                 \\
RS                     & (0,1,0) & 1            & 9           & 1    & 1E-04       & 3                                                                                 \\
SC                     & (0,2,1) & 10           & 0           & 1    & 3.16E-03       & 5                                                                                 \\
SP                     & (0,2,0) & 20           & 9           & 1    & 1E-04       & 5                                                                                 \\ \hline
\end{tabular}
}
\end{table}

\end{document}